\documentclass[prb,twocolumn,aps,showpacs,superscriptaddress,epsfig,longbibliography]{revtex4-2}

\usepackage{amssymb}
\usepackage{amsbsy}
\usepackage{amsmath}
\usepackage{float}
\usepackage{epsfig}
\usepackage{graphicx}
\usepackage{subfigure}
\usepackage[dvipsnames]{xcolor}

\begin{document}
\title{Extended critical phase in quasiperiodic quantum Hall systems}

\author{Jonas F.~Karcher}
\affiliation{{Pennsylvania State University, Department of Physics, University Park, Pennsylvania 16802, USA}}

\author{Romain Vasseur}
\affiliation{{Department of Physics, University of Massachusetts, Amherst, Massachusetts 01003, USA}}

\author{Sarang Gopalakrishnan}
\affiliation{{Princeton University, Department of Electrical and Computer Engineering, Princeton, NJ 08544, USA}}

\date{\today}

\begin{abstract}
    We consider the effects of quasiperiodic spatial modulation on the quantum Hall plateau transition, by analyzing the 
    Chalker-Coddington network model 
    with quasiperiodically modulated link phases. 
    In the conventional case (uncorrelated random phases), there is a critical point separating topologically distinct integer quantum Hall insulators. Surprisingly, the quasiperiodic version of the model supports an extended critical phase for some angles of modulation. We characterize this critical phase and the transitions between critical and insulating phases. For quasiperiodic potentials with two incommensurate wavelengths, the transitions we find are in a different universality class from the random transition. Upon adding more wavelengths they undergo a crossover to the uncorrelated random case. We expect our results to be relevant to the quantum Hall phases of twisted bilayer graphene or other Moir\'e systems with large unit cells.
\end{abstract}

\maketitle
\section{Introduction}

The integer quantum Hall (IQH) effect is the remarkably robust quantized Hall response of two-dimensional electron gases (2DEGs) subject to a strong external magnetic field. Disorder plays a crucial part in stabilizing plateaux of density with quantized Hall response~\cite{girvin1987quantum}: almost all the single-particle states in the IQH regime are Anderson localized, and moving the Fermi energy in a region of localized states does not change the response. Plateau transitions occur when the Fermi level crosses an extended state, leading to a jump in the quantized response. In the ``standard'' IQH scenario, with uncorrelated randomness, the plateau transition is a critical point, about which many open questions remain~\cite{PhysRevLett.61.1294,PhysRevLett.61.1297,PRUISKEN1984277,ZIRNBAUER2019458, karcher2021generalized, karcher2022generalized}. However, in many present-day realizations of IQH physics---such as graphene grown on a substrate, or twisted bilayer materials in a magnetic field---the dominant spatial modulations are not uncorrelated but quasiperiodic. The study of electronic states---and more generally wave propagation---in quasiperiodic media has been a topic of intense experimental interest~\cite{zilberberg2018, he2021moire, shimasaki2022anomalous, morison2022order, vaidya2022reentrant}. Wavefunctions in quasiperiodic media also undergo Anderson localization, but the nature of the localization transition is different from that in random systems. The best-studied example of a quasiperiodic potential the Aubry-Andr\'e model in one dimension~\cite{Aubry_1980}, which exhibits a transition from ballistic to localized states as the potential strength is tuned. (In contrast, random systems in one dimension are always in the localized phase~\cite{kramer1993localization}.) Recently, the localization transition in higher-dimensional or longer-range quasiperiodic systems has also been studied~\cite{devakul2017anderson, PhysRevB.96.054202, pixley2018weyl, fu2020magic, PhysRevB.99.054211, chou2020magic, szabo2020mixed, PhysRevB.106.184209, shimasaki2022anomalous}, but not for the symmetry class~\cite{altland1997nonstandard} corresponding to the plateau transition.  In addition to the ballistic and localized phases, some of these models have been shown to exhibit unusual intermediate phases, but these have not yet been classified. 

\begin{figure}[b]
	\centering 
	\includegraphics[width=.99\linewidth]{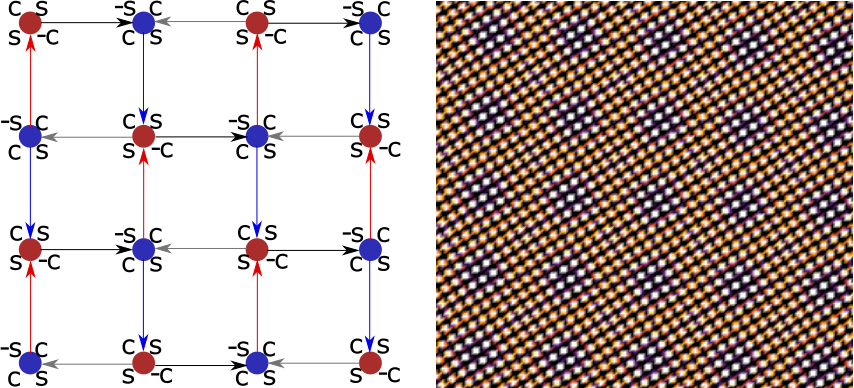}
 
	\caption{{\bf Network model.} \textit{Left panel:} Visualization of the scattering matrix $\mathcal{S}$ of Chalker-Coddington network model defined on a square lattice. The degrees of freedom live on the links, receive phases while propagating, can scatter into each other at the nodes weighted by $\pm s\equiv\pm \sin(\rho)$ or $ \pm c\equiv \pm \cos(\rho)$. At $\rho= \pi/4$, the quantum Hall transition occurs.  \textit{Right panel:} Quasiperiodic configuration ($\theta=\pi/6.2$) for the link phases. The scattering angles are tuned to criticality $\rho= \pi/4$ for the ED numerics. }
	\label{fig:scat}
\end{figure}

In this work we study the effects of quasiperiodic spatial modulations on the IQH plateau transition. Following the standard approach to the disordered case, we study Chalker-Coddington (CC) network models \cite{chalker1988percolation, kramer2005random} with quasiperiodically modulated link phases. In most of the paper, we consider the simplest type of quasiperiodic modulation, namely, the case in which the parameters in the network model are modulated with a single wavelength that is incommensurate with the underlying lattice structure. This case is also the most relevant to potential near-term experiments, e.g., on Moir\'e materials in magnetic fields~\cite{he2021moire}, graphene grown on a substrate, or ultracold atomic gases in synthetic magnetic fields~\cite{goldman2014light}. However, in order to understand how the quasiperiodic transition and the random one are related, we also consider systems with multiple incommensurate wavelengths. Our main results come from direct numerical calculations of the single-particle states, but our conclusions are also qualitatively supported by a real-space renormalization group treatment of a simplified model.

\subsection{Outline of this work}
In Sec.~\ref{sec:cc}, we introduce the Chalker-Coddington network we focus on and the scaling theory we use to probe its phase diagram (Sec.~\ref{sec:scaling}). Our main results are as follows. We show that the plateau transition in the quasiperiodic network model with two incommensurate wavevectors (``tones'') lies in a different universality class from the standard plateau transition in Sec.~\ref{sec:phase}. Indeed, the phase diagram of the network model is richer in the two-tone quasiperiodic case: instead of two insulating phases separated by a critical point, we find (in some parameter ranges, see Sec.~\ref{sec:ext}) a critical \emph{phase} between the two insulating phases. When this critical phase is present, the quantized Hall plateaux are separated by a metallic phase in which the Hall conductivity is not quantized. In addition, even for parameters that show a direct transition between two insulating phases, the critical exponents at this transition differ from the exponents in the standard plateau transition (see Sec. \ref{sec:dir}). Adding more tones induces a crossover to the random critical behavior, but not when the additional tones are sufficiently weak, which is discussed in Sec. \ref{sec:cross}. 

\section{Models and methods} 
Here we introduce the Chalker-Coddington network we focus on (\ref{sec:cc}) and the Ando model (\ref{sec:ando}) we consider for the crossover to randomness. In Sec. ~\ref{sec:scaling}, we introduce the scaling theory necessary to find the phase diagrams of our models.

\subsection{Quasiperiodic Chalker Coddington network}
\label{sec:cc}
The CC network model~\cite{chalker1988percolation}  is an effective model of electronic states near a plateau transition. It is a model of chiral degrees of freedom defined on the links of a square lattice, which scatter either left or right at the vertices of the lattice (Fig. \ref{fig:scat}). Each vertex hosts a unitary scattering matrix $\mathcal{S}$, with scattering weights $\pm s\equiv\pm \sin(\rho)$ or $ \pm c\equiv \pm \cos(\rho)$. The network model can be interpreted as a Floquet unitary evolution~\cite{PhysRevLett.125.086601}. {Its single particle spectrum lies in the unit circle of the complex plane and the eigenvalues $e^{i\omega}$ can be labeled by the real quasienergy $\omega$.}
Deep in the two localized phases, $|c| \approx 1$ and $|s| \approx 1$, the eigenstates of the network model are localized on single plaquettes, and rotate clockwise (counterclockwise) for $|c| \approx 1$ ($|s| \approx 1)$. For random potentials, the plateau transition occurs at the self-dual point $\rho_s = \pi/4$. Knowing the location of the self-dual point allows us to eliminate a class of finite size effects related to the uncertainty of the position of the critical point. We denote the distance to this point by $d= \rho - \rho_s$. {In the quasiperiodically modulated model there still is symmetry around the self-dual point, in other words $\rho$ can be exchanged for $\rho_s-\rho$ (and $\rho \rightarrow \pi+\rho$).}

\begin{figure*}[t]
	\centering
 	\includegraphics[width = \textwidth]{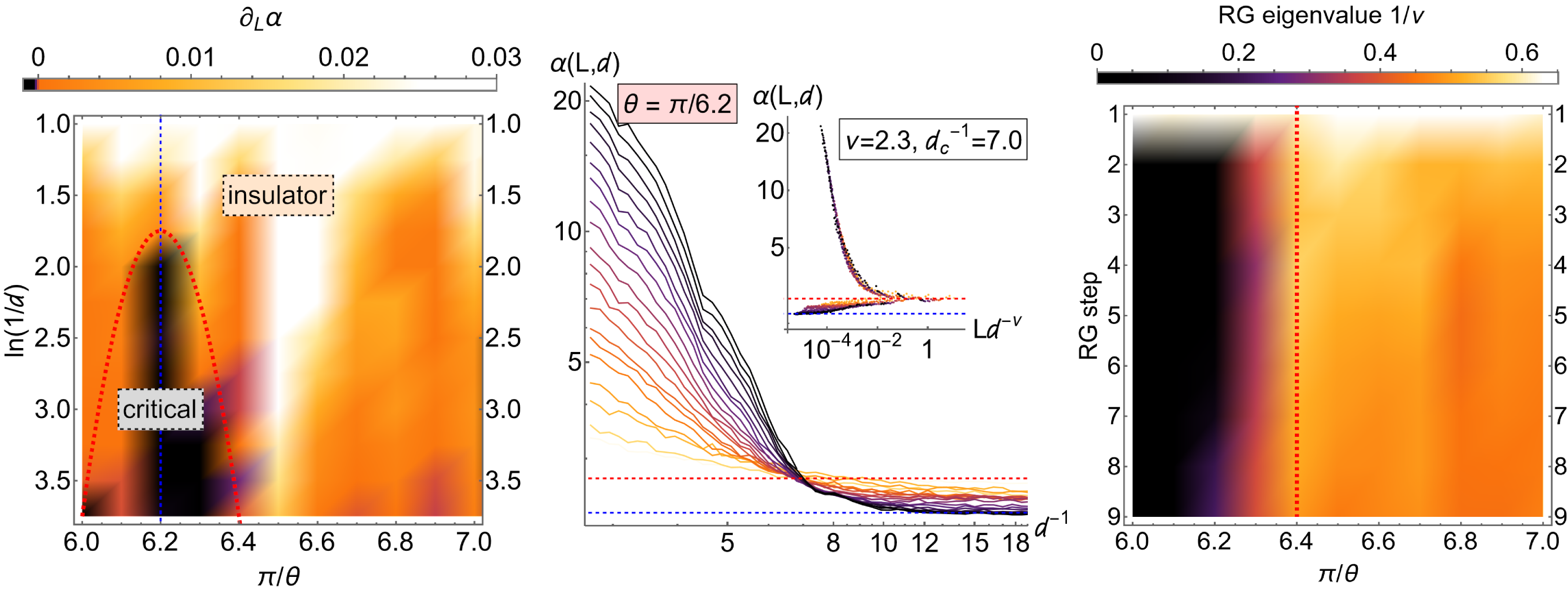}
	\caption{{\bf Phase diagram and stability of the extended critical phase}. \textit{Left panel:} Derivative $\partial_L\alpha(L,d)$ of the finite size scaling function for large system sizes. The red dashed line showing the approximate phase boundary is a guide to the eye.
	\textit{Middle panel:} Quasiperiodic angle $\theta=\pi/6.2$ with extended critical phase. The transition point can be estimated to be at 
	{$\ln d^{-1} = 1.94$}. We use $N=500$ configurations of the phases $\gamma_{1,2}$ to determine $\alpha(L,d)$ for each data point. Red and blue dashed lines mark the limiting value of $\alpha$ at the critical point (crossing) and in the extended critical phase.
	\textit{Right panel:} The first few iterations of a toy model real space renormalization group applied to the quasiperiodic network model. The flow of detunings $d$ from the self-dual point is associated with the RG eigenvalue $\nu^{-1}$ -- the inverse correlation length exponent. For $\theta<\pi/6.4$, there is a region where small but finite detunings $d$ are exactly marginal.}
	\label{fig:mf_phase}
\end{figure*}

We choose link phases $\phi(r)$ at each link $r$ defined over the quasiperiodic function {(following Refs. \cite{szabo2020mixed, devakul2017anderson})}:
\begin{align}
    \phi(r) = 2\pi\sum_{i=1,2} \cos\left(\sum_jA_{ij}r_j + \gamma_i\right), \nonumber\\
    A_{11} = A_{22} = \varphi\cos\theta, A_{12} = - A_{21} = \varphi\sin\theta. \label{eq:qp},
\end{align}
here $\varphi=(1-\sqrt{5})/2$ is the golden ratio and $\gamma_{1,2}$ are phases. We show an example  configuration ($\theta=\pi/6.2$) for the link phases, it exhibits a superlattice moire pattern with an emergent length scale of several lattice constants. {At $\pi/5$, the twist $\cos(\pi/5)$ becomes proportional to the golden ratio and this makes the phases commensurate with the lattice in one direction. Further, at $\pi/\theta$ with large $\theta$, there is basically no rotation ($\cos(\pi/\theta)\approx 1$). These endpoints determine the range of $\theta$ investigated in this work.} In order to avoid the need for rational approximants of the angles $\theta$ and the golden ratio matching the system size (see discussions in Refs.~\onlinecite{devakul2017anderson, szabo2020mixed}), we choose open boundary conditions. This comes at the expense of having to omit values of the wavefunction close to the boundary in the multifractal analysis.

\subsection{Quasiperiodic Ando model}
\label{sec:ando}
We are further considering the Ando model in the symplectic class:
\begin{align}
	&H= \sum_{r,\sigma} \epsilon_r c^\dagger_{r,\sigma} c_{r,\sigma}\nonumber\\ &+\sum_{\sigma\sigma'} \left[c^\dagger_{r,\sigma} T^{\hat{x}}_{\sigma\sigma'} c_{r+\hat{x},\sigma'} + c^\dagger_{r,\sigma} T^{\hat{y}}_{\sigma\sigma'} c_{r+\hat{y},\sigma'} + h.c.\right],\nonumber\\
	&T^{\hat{x}} = \begin{pmatrix}
		\frac{\sqrt{3}}{2} & \frac{i}{2}  \\
		\frac{i}{2} & \frac{\sqrt{3}}{2}
	\end{pmatrix} , \quad\quad T^{\hat{y}} = \begin{pmatrix}
		\frac{\sqrt{3}}{2} & \frac{1}{2}  \\
		-\frac{1}{2} & \frac{\sqrt{3}}{2}
	\end{pmatrix}.
\end{align}
Quasirandomness enters through the potential     $\epsilon_r$. The phase diagram of this model has been explored in Ref. \cite{devakul2017anderson}. We back up our claims on crossing over to the random fixed point with this model in Sec. \ref{sec:rand_ando}.

With this in mind, we define $\epsilon_r = V (w \phi_1(r) + (1-w)\phi_2(r))$ with two functions $\phi_k(r)$ defined analogous to Eq.~\eqref{eq:qp}. The parameter $w\in [0,1]$ controls the relative strength of the different tones. For the uncorrelated random case to compare to, we choose $\epsilon_r \in [-V/2,V/2]$ independently and uniformly.

\subsection{Scaling theory and Observables}
\label{sec:scaling}
At Anderson transitions, there is an infinite continuum of critical exponents, the multifractal spectrum. {When the potential is spatially uncorrelated, the multifractal spectrum can be extracted from the scaling of $q$-th local density of states (LDOS) moments 
\begin{align}
\left\langle \rho(\omega, r)^q\right\rangle_\beta \sim \left\langle\rho(\omega)^q\right\rangle_\beta\left\langle \; |\psi(r)|^{2q} \right\rangle_\beta \sim L^{-2q  - x_q}
\end{align}
in systems where the density of states $\rho(\omega)$ does not scale with the system size. This defines a scaling dimension $x_q$ for each $q$, the multifractal spectrum.}

For correlated potentials, it is necessary to average the wavefunctions over boxes larger than the correlation volume first before analyzing the moments:
\begin{align}
	\mu_i &= \int_{B_i} d^2r \; |\psi(r)|^2, &  \left\langle\ln \mu_i \right\rangle_\beta &= \alpha \ln \left(\dfrac{b}{L}\right).
\end{align}
When one chooses the box size $b$ to scale like the system size, say $b=L/12$, then the logarithmic average 
\begin{align}
	\alpha = \dfrac{1}{N_B}\sum_i \langle \ln(\mu_i) \rangle_\beta, \label{eq:fss}
\end{align}
{is a quantity suitable to a finite-size scaling (FSS) study to diagnose criticality \cite{devakul2017anderson}. Here division by $N_B$ (the number of boxes) counters the sum over all boxes. Close to criticality (measured by a parameter $d$) and for large systems, this quantity obeys the scaling form $\alpha(L, d) = \alpha(L(d-d_c)^{-\nu})$, ignoring leading irrelevant corrections~\cite{slevin1999corrections}.} Since we have open boundary conditions, we need to drop the layer of $\mu_i$ adjacent to the boundary. 

\section{Phase diagram of QP Chalker Coddington network}
\label{sec:phase}
In this section, we investigate the extended critical phase appearing in the quasiperiodic quantum Hall transition in Sec.~\ref{sec:ext} and contrast that with the direct transition in Sec.~\ref{sec:dir}. Further we look into the dynamical properties of the network interpreted as a Floquet circuit in Sec. \ref{sec:dyn}.

\subsection{Extended critical phase} 
\label{sec:ext}
We find the phase diagram of the quasiperiodic network model using this scaling theory approach. Remarkably, there is an extended critical phase not present in the random $U(1)$ CC network around the self-dual line.  
In the left panel of Fig.~\ref{fig:mf_phase}, we show the derivative $\partial_L\alpha(L,d)$ of the finite size scaling function for large system sizes. We identify the different phases as follows: as $L \to \infty$, $\partial_L\alpha(L,d)>0$ in the insulator, $\partial_L\alpha(L,d)<0$ in the metal, and at critical points $\partial_L\alpha(L,d)=0$. At the critical point, the FSS amplitude $\alpha$ is constant and characterizes the fractality of wavefunctions. In 2D, it can be challenging to distinguish bad metals from true asymptotic critical points, since the corresponding RG flows in the uncorrelated case can be as slow as logarithmic in $L$ \cite{karcher2022generalizedMIT}. In the IQH symmetry class A such effects are conventionally not present at strong randomness and the criticality we observe in the finite-size systems here is a genuine feature of the quasirandomness whether or not it persists to $L\rightarrow\infty$.

For certain angles, we perform a more detailed finite size scaling analysis with higher resolution in $d$. For example for the quasiperiodic angle $\theta=\pi/6.2$ that supports the extended critical phase, we can estimate the transition between critical and insulating phases to be at {$\ln d^{-1} = 1.94$}. The critical phase we find seems to have a constant value of $\alpha$, which differs from the value of $\alpha$ at the critical endpoint at {$\ln d^{-1} = 1.94$}. Our data is shown in the middle panel of Fig. \ref{fig:mf_phase}.

We support our finding of a critical phase by studying a toy model for real-space renormalization group (RG) applied to the quasiperiodic network model~\cite{kramer2005random, galstyan1997localization, arovas1997real}. We compute the flow of detunings $d$ from the self-dual point; the RG eigenvalue $\nu^{-1}$ of $d$ is the correlation length exponent. For $\theta<\pi/6.4$, there is a region where small but finite detunings $d$ are exactly marginal (stable extended critical phase, black color in plot Fig.~\ref{fig:mf_phase}(c)). This qualitatively matches the exact diagonalization result. (Note that the real-space renormalization group is not asymptotically exact for this model, so one does not expect quantitative agreement with numerics.)

\begin{figure}
	\centering
 \includegraphics[width = \linewidth]{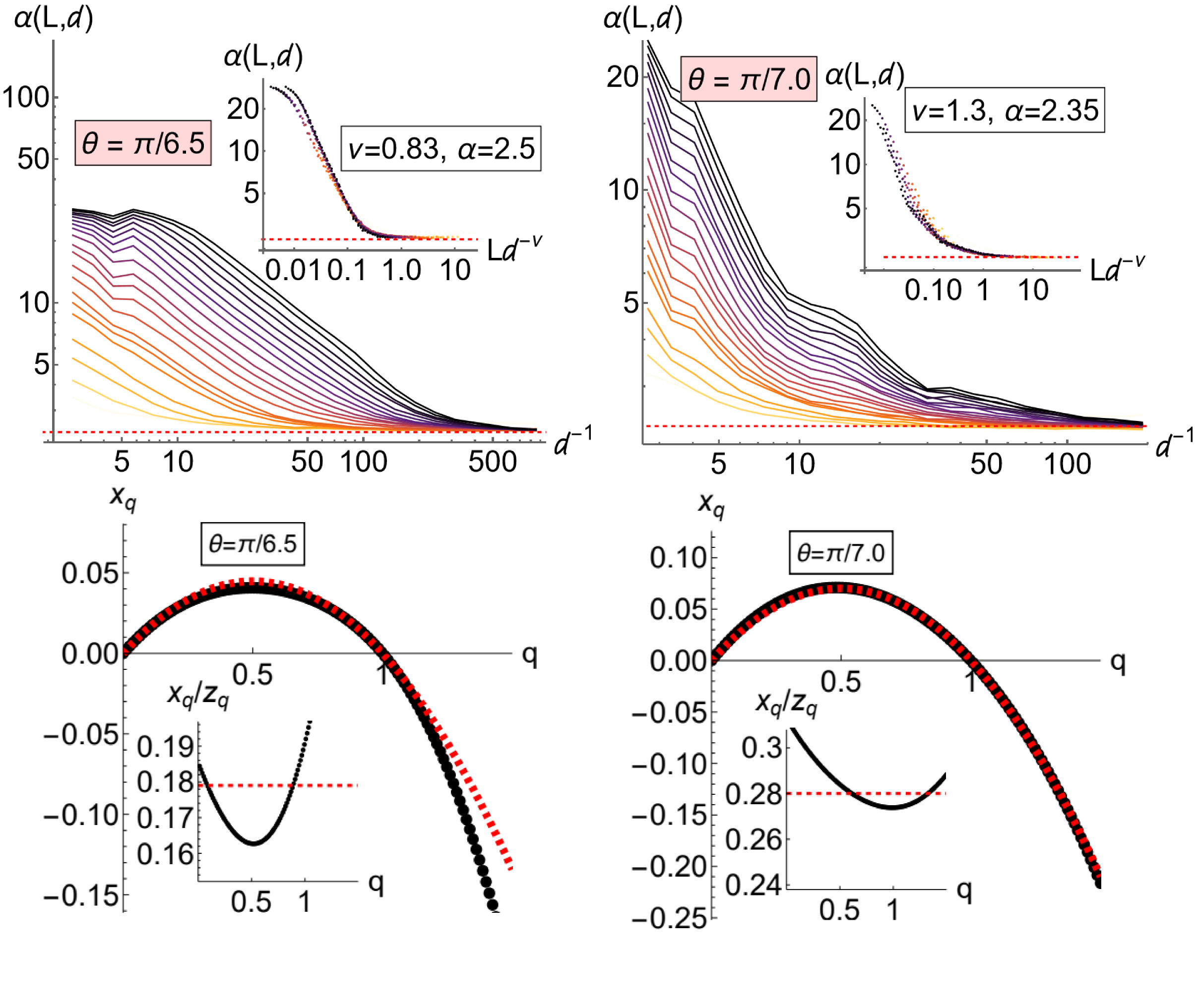}
	\caption{{\bf Localization length exponent.} Determination of the universal localization length exponent $\nu$ using a finite size function $\alpha(L, d)$ defined in Eq. \eqref{eq:fss}. Different colors represent different system sizes. In the insets finite size collapses are shown. \textit{Left panel:} Quasiperiodic angle $\theta=\pi/6.5$ with fast flow away from the critical point $\nu\approx 0.8\pm 0.05$. \textit{Right panel:} Quasiperiodic angle $\theta=\pi/7.0$ with slow flow away from the critical point $\nu\approx 1.3\pm 0.1$. 
	\textit{Lower panels:} Multifractal scaling dimensions $x_q$ in comparison to parabolic form $\beta z_q$ (dashed red line). }
\label{fig:mf_fss}
\end{figure}

\subsection{Direct transition}
\label{sec:dir}
We now turn to values of $\theta$ for which a direct transition between two insulating phases persists in the quasiperiodic case. For the determination of the universal localization length exponent $\nu$, we use the finite size function $\alpha(L, d)$ defined in Eq.~\eqref{eq:fss}.  In Fig.~\ref{fig:mf_fss}, we show results for two different quasiperiodic potentials with a direct transition. 
{For the determination of the data points, we use linear system sizes $L=12$, 24, 36, 48, 72, 84, 96, 120, 144, 168, 192, 216, 240, 288, 336, 384, 480, 540, 600 with $N=500$ configurations of the phases $\gamma_{1,2}$ each. The statistical errors are of the order of the point size. For the fits determining the critical exponent, we only take $L\geq 96$ into account.} For the quasiperiodic angle $\theta=\pi/6.5$ shown in the left panel of Fig.~\ref{fig:mf_fss} we find fast flow away from the critical point with the exponent $\nu\approx 0.8\pm 0.05$. In the right panel, we show the angle $\theta=\pi/7.0$ with slow flow away from the critical point $\nu\approx 1.3\pm 0.1$. Given the long length scales involved in the moire patterns, we cannot exclude a scenario where these exponents will eventually flow to $\nu=1$ in the thermodynamic limit. Recall that, by the Harris criterion, $\nu \geq 1$ in two dimensions would imply the stability of the two-tone quasiperiodic critical point in the presence of additional weak uncorrelated randomness. With the available system sizes we are unable to definitively address this question, but it remains an interesting one for future work.

\subsection{Non-universal multifractal spectra} In the conventional uncorrelated random case, the IQH multifractal spectrum is a universal property of the critical point. 
{The multifractal spectrum is defined by the scaling of the $q$-th participation ratios:
	\begin{align}
		\mathcal{P}_q\equiv \frac{1}{N_B}\left\langle \sum_i (\mu_i)^q \right\rangle \sim (b/L)^{2q+x_q} \label{eq:qpr}.
	\end{align}
	These anomalous dimensions encode the information about the fractal dimensions $f(\alpha)$ of the sets where the wavefunction scales as $L^{-\alpha}$, more precisely $(q,x_q)$ is the Legendre transform of $(\alpha, f(\alpha))$ the technical mathematical construction is reviewed in Ref. \cite{Evers2008}. Typically the spectrum is approximately parabolic $x_q\approx \beta q(1-q) \equiv \beta z_q$. }

In the quasiperiodic case we find, once again, that these spectra behave quite differently for the direct transitions at $\theta = \pi/6.5$ and at $\theta = \pi/7$. 
For the determination of multifractal spectra, we use $N=10^4$ configurations of the phases $\gamma_{1,2}$. In the lower panels of Fig.~\ref{fig:mf_fss} we show the multifractal spectra $x_{q}$ for these two cases. The black dots are data points from a fit of the exponent over system sizes 
{$L=96, 144, 216, 336, 480, 600$. For $\theta=\pi/6.5$, we observe $\beta \approx 0.18$ and approximate parabolicity. The spectrum is not universal because for $\theta=\pi/7.0$, we observe a different curvature $\beta \approx 0.28$. In the appendix~\ref{app:multi}, we analyze different points in the phase diagram that show more irregular behavior of the scaling dimensions $x_q$. }

\begin{figure}[t!]
	\centering
	\includegraphics[width = .9\linewidth]{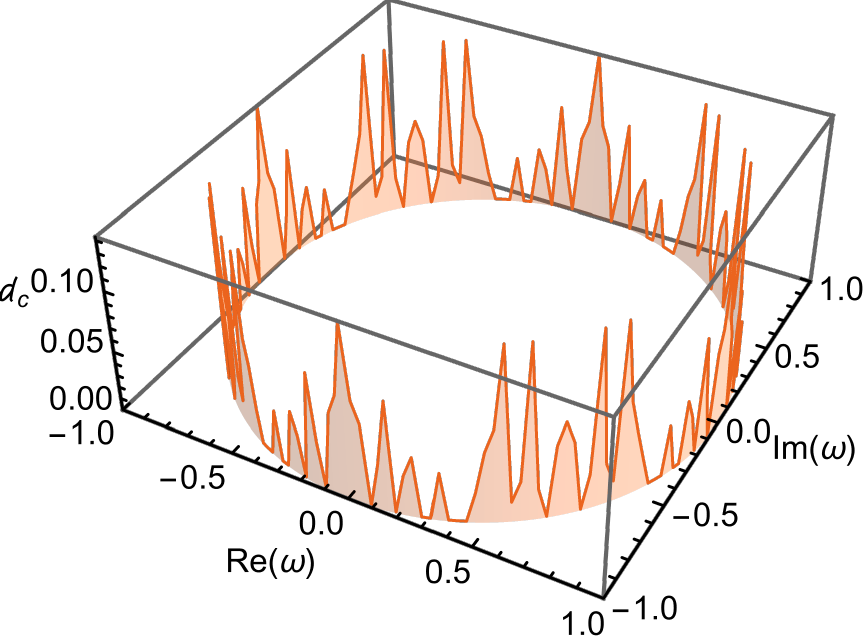}
	\caption{{\bf Quasienergy spectrum.} Quasienergy dependence of the transition $d_c$ to the extended critical phase ($\theta=\pi/6.2$). For a direct transition between the topological phases, we put $d_c=0$. Re/Im$\omega$ plane is the complex quasienergy plane, the spectrum of the unitary Chalker-Coddington scattering matrix lies on the unit circle. The $d_c$ axis shows the maximum extent of the critical phase at a given quasienergy $\omega$ on the unit circle. }
	\label{fig:metal_en}
\end{figure}

\subsection{Dynamical properties}
\label{sec:dyn}
Here we study the quasienergy dependence in the network model and the corresponding implications for its dynamical properties in several observables.

\subsubsection{Quasienergy dependence of phases}
If one regards the network model as a Floquet system~\cite{PhysRevLett.125.086601}, the eigenstates of the model are labeled by a \emph{quasienergy} $\omega$. The quantum Hall problem nominally corresponds to $\omega = 0$, although there are indirect ways to extract transport properties from $\omega$-dependence~\cite{klesse1997spectral}.An important point to note is that in the uncorrelated random problem wavefunction statistics and level repulsion behavior of all $\omega$ is the same at distribution level. 

\begin{figure*}[t]
	\centering
	\includegraphics[width = .32\textwidth]{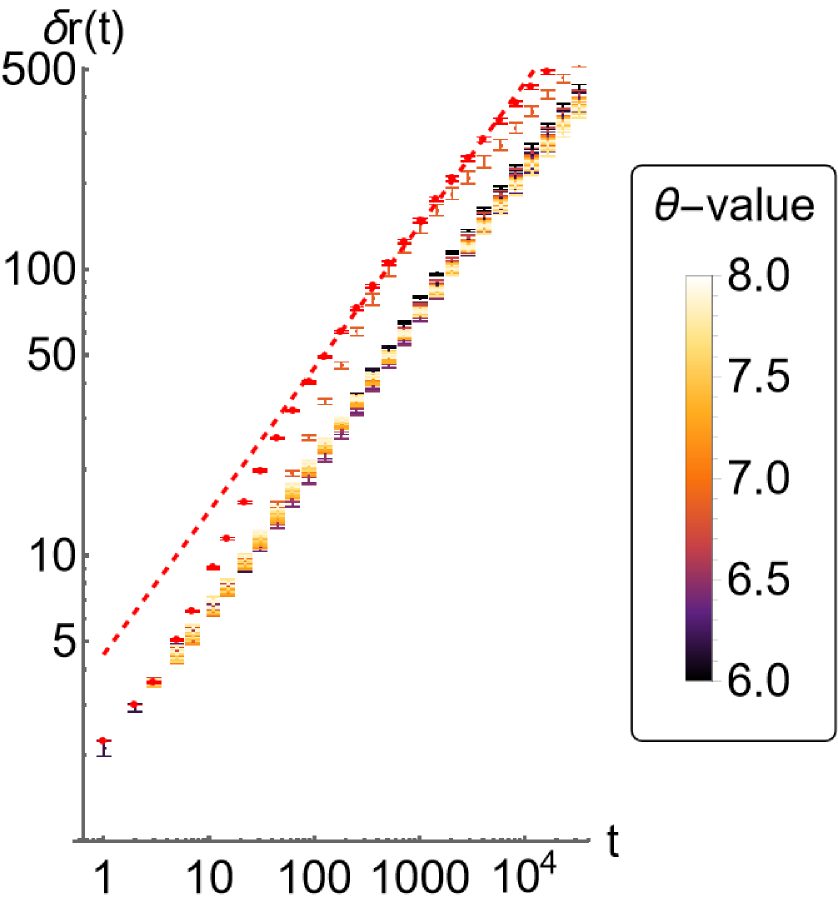}
	\includegraphics[width = .32\textwidth]{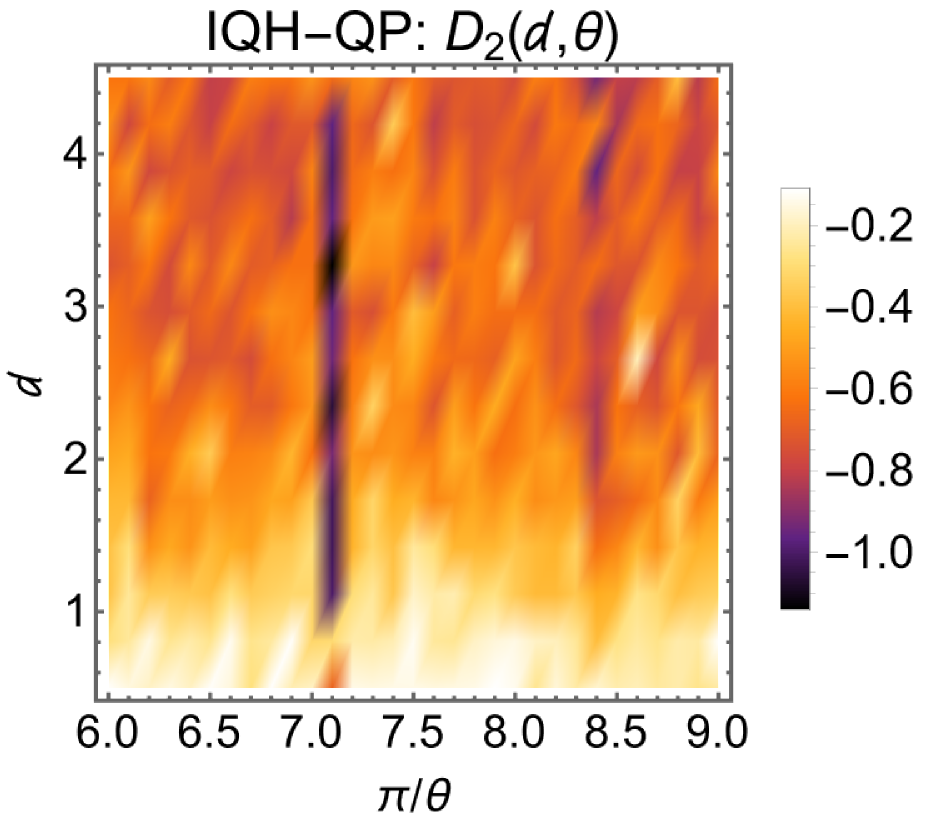}
	\includegraphics[width = .32\textwidth]{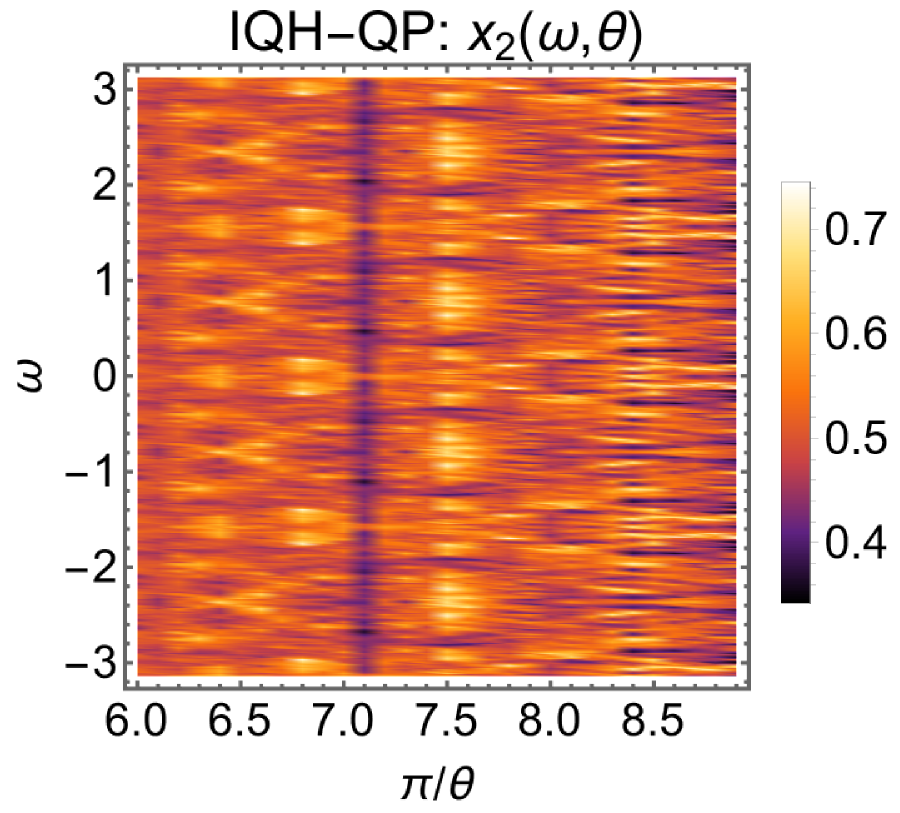}
	\caption{{\bf Dynamical properties of the quasiperiodic $U(1)$ CCN.} Comparison of return probability to IPR scaling for a range of $\theta=\pi/6.0$ to $\theta=\pi/9.0$. \textit{Left panel:} spread $\delta {r} (t)$ as function of time $t$ for random (red) vs quasiperiodic phases (sunset) at the self dual point \textit{Middle panel:} the extended critical phase for $\theta=\pi/6.0$ leaves a visible imprint here \textit{Right panel:} the IPR shows a complex (probably fractal) dependency across the quasienergy spectrum.}
	\label{fig:dyn}
\end{figure*}

In our quasiperiodic problem, we now consider how the spectrum at general $\omega$ evolves as one tunes $d$ (Fig.~\ref{fig:metal_en}), for an angle $\theta=\pi/6.2$ where an intermediate metallic phase exists at $\omega = 0$. The finite-$\omega$ spectrum can be probed, for example, in atomic or optical systems that directly realize the network model. Just like the $\theta$-dependence, we find that the $\omega$-dependence is also highly irregular, with multiple lobes of intermediate extended criticality separated by quasienergies for which one has a direct transition.  This quasienergy dependence has interesting implications for the dynamics of the network, which is studied in the following paragraphs. 

\subsubsection{Return probability}
A well-established observable in the context of dynamics is the return probability:
\begin{align}
	\langle P (t)\rangle &= \left\langle |\langle \psi(t)|\psi(0)\rangle|^2\right\rangle, & \psi(t) &= \sum_n \langle\psi_n|\psi(0)\rangle e^{i n t}|\psi_n\rangle.
\end{align}
Results for time evolution simulations are in left panel of Fig.~\ref{fig:dyn}. We show $D_2(d, \theta)$ as function of quasiperiodic angle $\theta$ and distance from critical point $d$. There is a pronounced dependence on both of these quantities.

\subsubsection{Wave packet spread}
We study the time evolution of an initially localized  wavepacket $\psi(0)$ in our quasiperiodic system. The complex quasienergy landscape (see Fig.~\ref{fig:metal_en}) adds features compared to the uncorrelated random case where every level behaves statistically the same. 

For an initially fully localized wavepacket $\langle {\bf r}|\psi(0)\rangle = \delta_{{\bf r},{\bf r}_0}$ at ${\bf r}_0=0$ all quasienergies and momenta are involved. The width $\delta {\bf r} (t) = {\bf r}(t)-{\bf r}_0$ usually grows diffusively at criticality:
\begin{align}    
	\langle \delta r^2 (t)\rangle = 
	\langle \psi(t)| ({\bf r}-{\bf r}_0)^2 |\psi(t)\rangle \sim D t.
\end{align}
In Fig. \ref{fig:dyn}, we show the spread $\delta {r} (t)$ as function of time $t$ for the random CCN compared to one with quasiperiodic phases at the self dual point. Diffusion is present irrespective of the quasiperiodic parameter and even the numerical value of the diffusion constant is quite insensitive to the angle $\theta$.

\subsubsection{Inverse participation ratio}
We investigate quasienergy dependence of the second inverse participation ratio (IPR):
\begin{align}    
	\langle |\psi^2_n({\bf r})|^2\rangle &\sim L^{-x_2(\epsilon_n)}.
\end{align}
For the uncorrelated random U(1) CCN the wavefunctions at all quasienergies obey the same statistical properties (see discussion in Refs.~\cite{kramer2005random, klesse1997spectral}), so the multifractal exponent $x_2$ and the wavepacket spread $D_2$ are trivially related:
\begin{align}
	t ^{-D_2} \sim \langle P (t)\rangle  \sim \sum_n\int\mathrm{d} {\bf r} \;\left\langle|\psi_n({\bf r})\psi_n({\bf r})|^2\right\rangle \sim L^{-2x_2},
\end{align}
since each term in the sum scales with the same power. In the quasiperiodic case, there is a complicated dependence of $x_2(\theta,\omega)$ in both quasiperiodic angle $\theta$ and quasienergy $\omega$. We show the results in right panel of Fig.~\ref{fig:dyn}. Nevertheless there are clear correlations of $x_2(\theta,\omega)$ to $D_2(d\rightarrow 0, \theta)$ (compare the upper region in left panel and right panel of Fig.~\ref{fig:dyn}).

\begin{figure}[t]
	\centering
	\includegraphics[width = .7\linewidth]{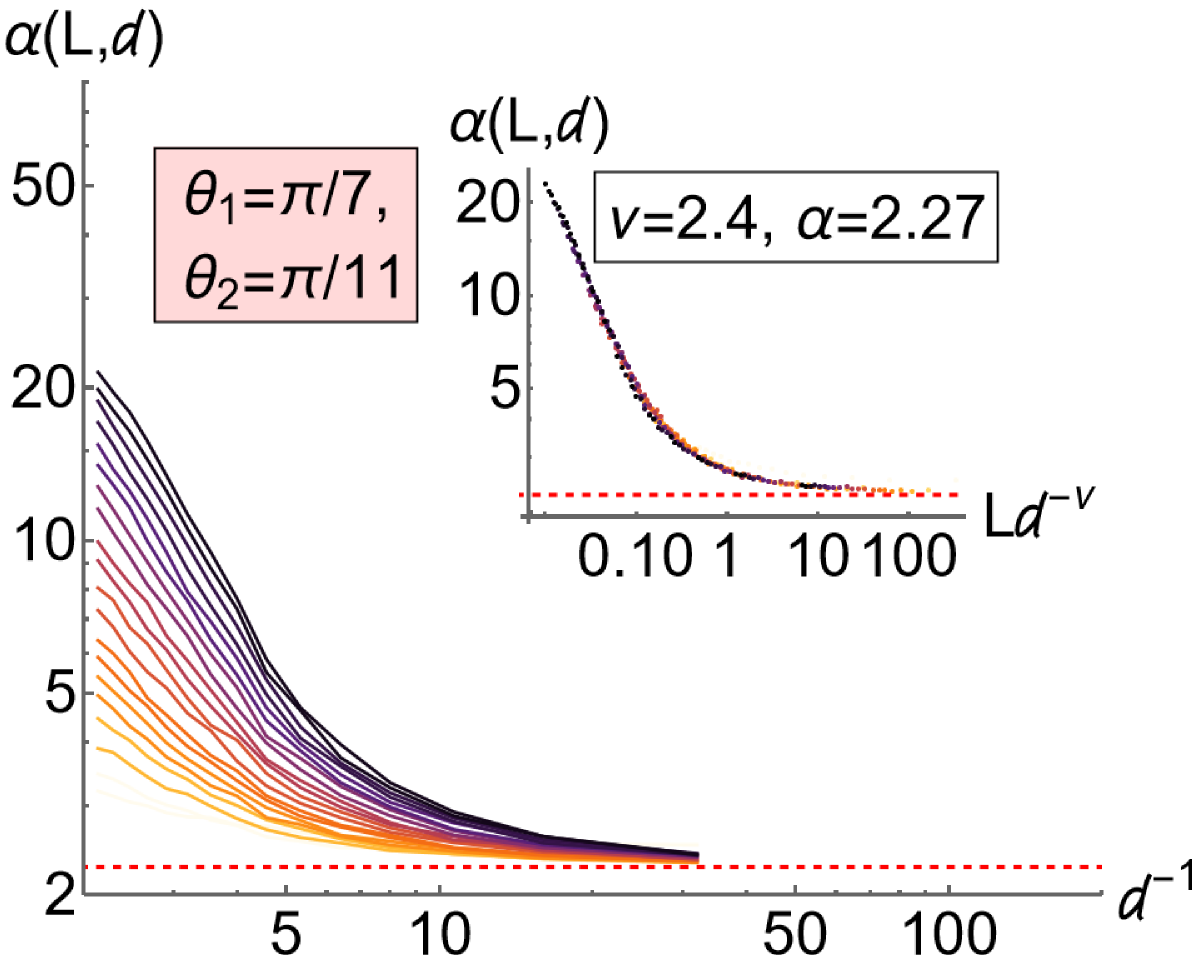}
	\caption{{\bf Crossover to random behavior for multiple tones.} The value of $\nu$ crosses over the the universal value $\nu\approx 2.4$ of the $U(1)$ CCN, when two quasiperiodic potentials with $\theta=\pi/7$ and $\theta=\pi/11$ are combined. The plot is analogous to the upper panels of Fig. \ref{fig:mf_fss}.}
\label{fig:crossover_iqh}
\end{figure}

\section{Crossover to random behavior}
\label{sec:cross}
The plateau transitions we found in the two-tone quasiperiodic model are strikingly different from the random case. We now discuss the crossovers that occur when one modifies the model to make it more like the random one, by adding more Fourier components (i.e., more tones) to the spatial modulation in Eq.~\eqref{eq:qp}. For simplicity we consider adding one additional incommensurate wavelength. 
\subsection{QP Chalker Coddington network}
When the two quasiperiodic modulations are of the same strength, we find that the plateau transition is direct, with no intermediate critical phase, and the critical exponent is numerically close to the IQH transition exponent $\nu_{\rm IQH}\approx 2.4$ \cite{Evers2008} (see Fig.~\ref{fig:crossover_iqh}). {At this point a comment is in order: high precision transfer matrix simulations of very long ($L=10^9$) strips obtain a different value $\nu_{\rm IQH}\approx 2.59$ \cite{slevin2009critical}. The sizes considered there exceed the typical unit cell size of QH Moire systems this  work was motivated by. } In Appendix \ref{app:transfer}, we present further supporting evidence for the conjecture that the presence of multiple tones leads to a crossover to the conventional uncorrelated random critical point. Interestingly, the crossover to the random critical point does not seem to occur when the second modulation is sufficiently weak: rather, the critical properties abruptly jump at some value of this modulation.  {The corresponding analysis is shown in Fig.~\ref{fig:crossover} of the Appendix \ref{app:transfer}.  
	
\subsection{QP Ando model}
\label{sec:rand_ando}
In order to further support our claim for the crossover to randomness, we demonstrate this effect also occurs in a different model and universality class, the Ando model introduced in Sec. \ref{sec:ando}. 
	
We determinate the universal localization length exponent $\nu$ using the finite size function $\alpha(L, d)$ defined in Eq.~\eqref{eq:fss}. Here the distance to the critical point $V_c$ is defined in terms of the strength of the potential $d = (V-V_c)/V_c$ . We use linear system sizes $L=36, 48, 72, 84, 96, 120, 144, 168, 192, 216$ and $N=5000$ phase configurations. The result is shown in Fig.~\ref{fig:crossover_ando}. Different colors represent different system sizes. In the insets finite size collapses are shown. We compare quasiperiodic with a single tone $\theta=\pi/8$ to quasiperiodic with $\theta=\pi/8$ and a second tone $\theta=\pi/6$ with equal strength and the random uncorrelated potential case. For the one tone case, the result agrees very well with the findings of Ref.~\cite{devakul2017anderson}. In the two tone case, the critical exponent and FSS amplitude approach the ordinary Anderson transition (AT) values exhibited at the random uncorrelated symplectic metal insulator transition, thus supporting our crossover claim also in this symmetry class. 
	
The spin degree of freedom enlargens the required matrix representation size and makes it computationally challenging to reach the necessary system sizes. We therefore leave a transfer matrix analysis of this model to future work.
	
	\begin{figure*}
		\centering
		\includegraphics[width = .3\textwidth]{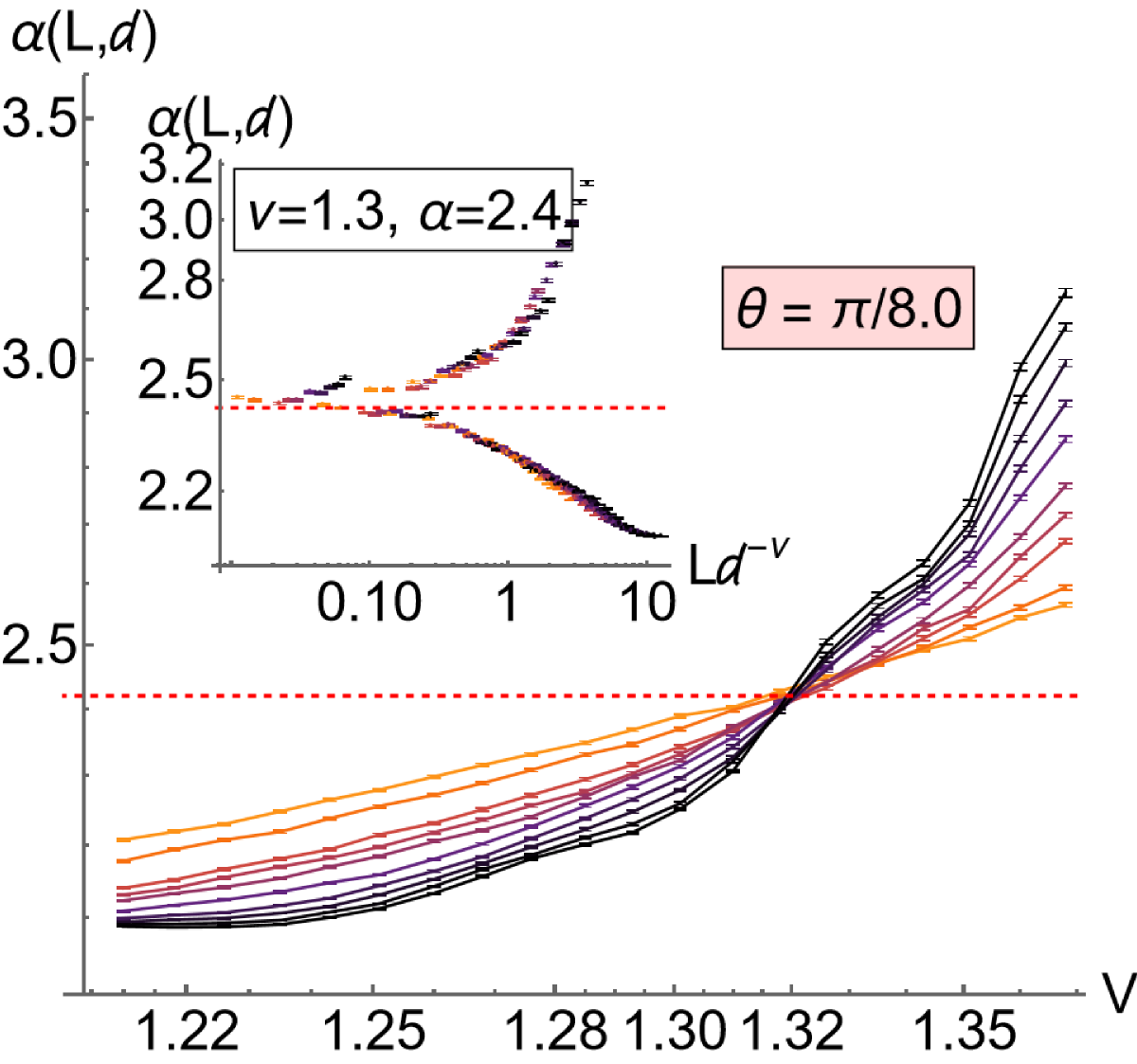}
		\includegraphics[width = .3\textwidth]{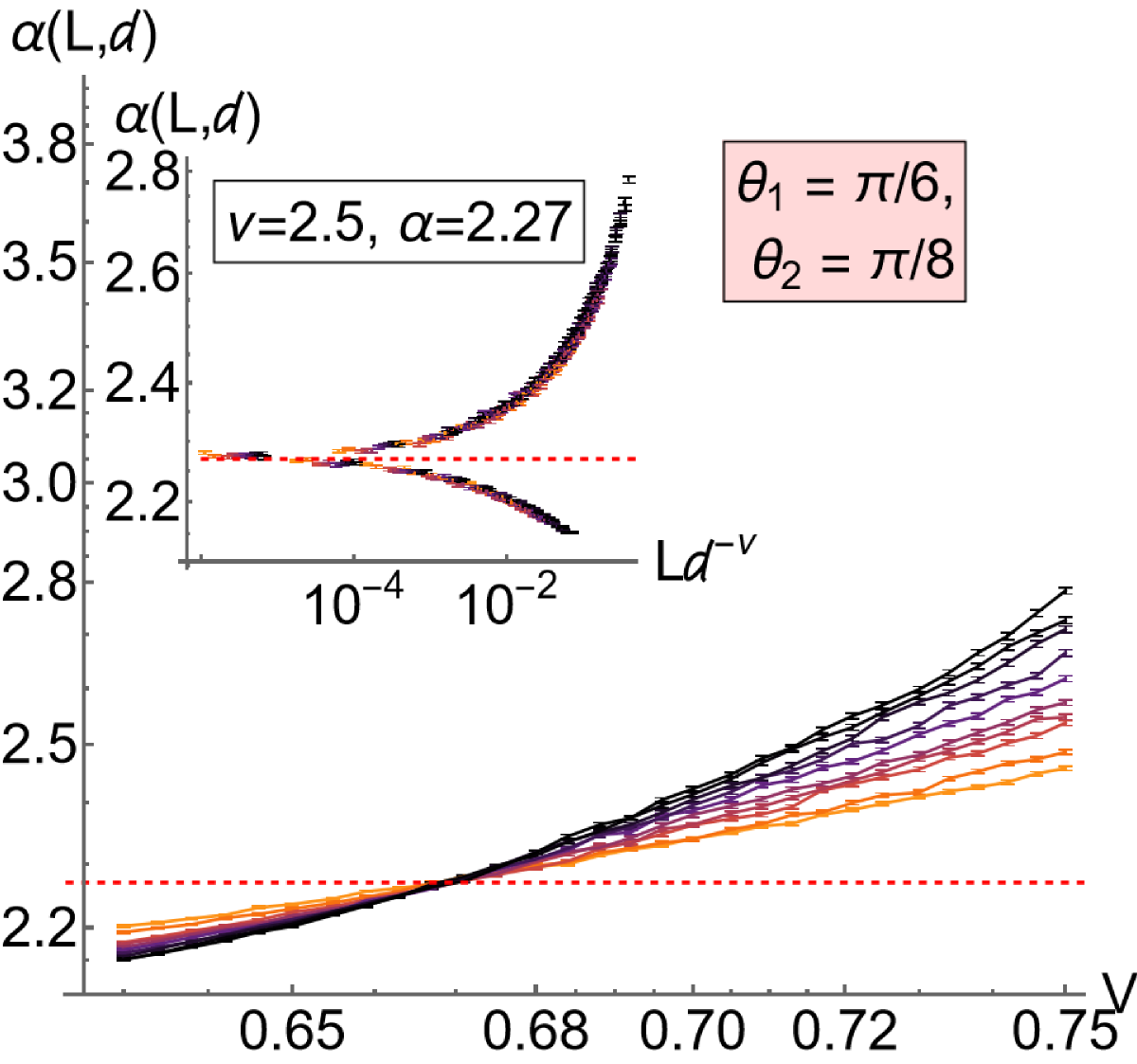}
		\includegraphics[width = .3\textwidth]{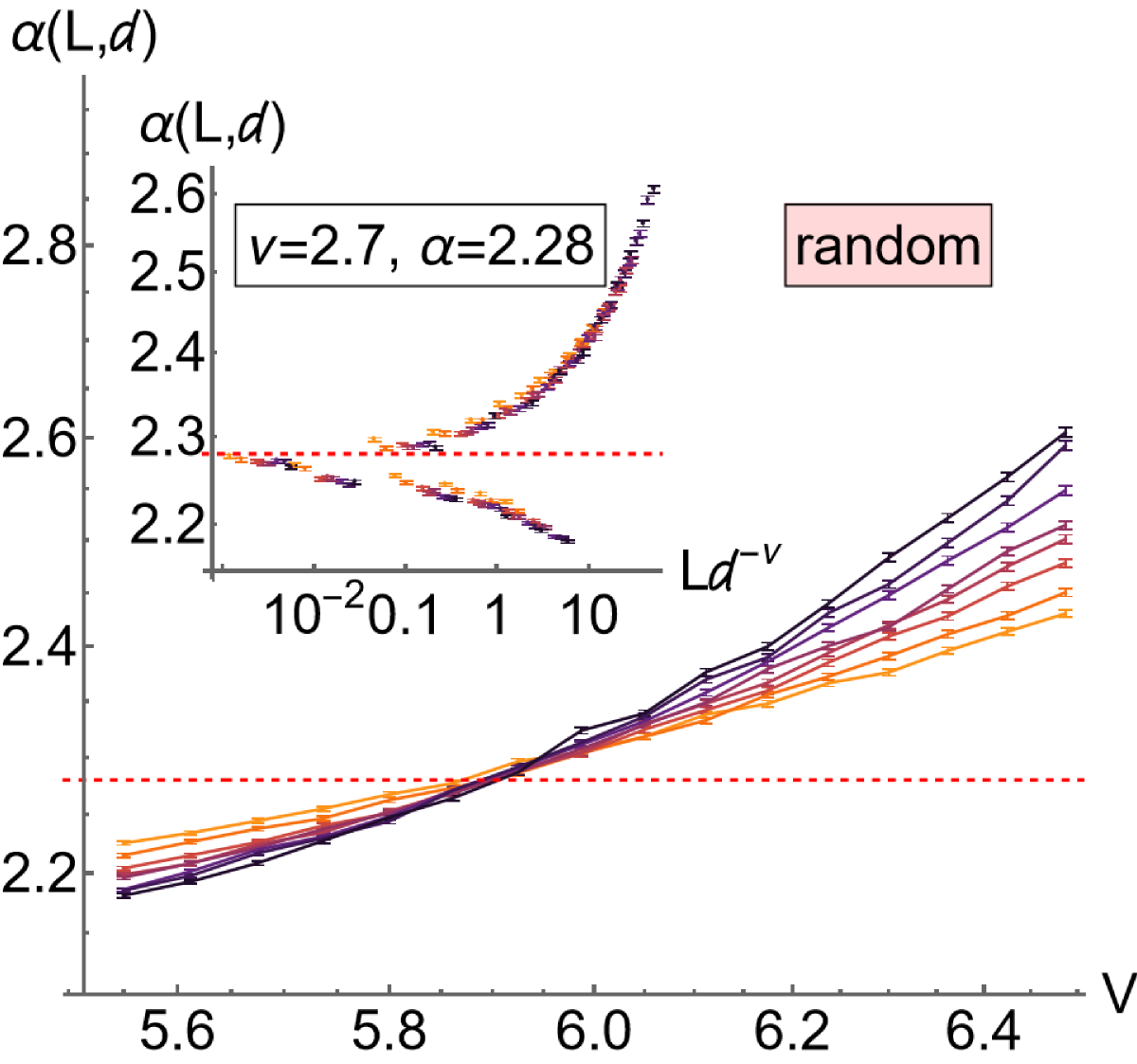}
		\caption{{\bf Finite size scaling analysis in the Ando model.} Determination of the universal localization length exponent $\nu$ using a finite size function $\alpha(L, d)$ defined in Eq.~\eqref{eq:fss}, in the Ando model where $w$ is the potential strength. Different colors represent different system sizes. In the insets finite size collapses are shown.
			\textit{Left panel:} Quasiperiodic with a single $\theta=\pi/8$. The result agrees very well with the findings of Ref.~\cite{devakul2017anderson}.
			\textit{Middle panel:} Quasiperiodic with $\theta=\pi/8$ and a second tone $\theta=\pi/6$ with equal strength. The critical exponent and FSS amplitude approach the ordinary AT values.
			\textit{Right panel:} Random uncorrelated potential, the model is the conventional Ando model \cite{Evers2008} in this limit. }
		\label{fig:crossover_ando}
	\end{figure*}

\section{Discussion} We studied the U$(1)$ Chalker-Coddington model with quasiperiodically modulated link phases. The critical properties of this model are distinct from the uncorrelated random version, which features (topological) insulator phases separated by critical points that can only be reached by fine tuning the energy or magnetic field. In the quasiperiodic case, the nature of the phase diagram is sensitive to the angle $\theta$ between the underlying lattice and the superimposed quasiperiodic modulation. For a range of $\theta$ we find  a critical phase between the two insulators. In the quantum Hall context, this means that quantized Hall plateaux are separated by a regime with nonvanishing longitudinal conductivity and non-quantized Hall response. For other values of $\theta$ we find a direct plateau transition; the associated critical exponent is clearly incompatible with the random case. The critical exponent $\nu$ that we find appears nonuniversal and $\theta$-dependent, ranging from $\nu = 0.80(5)$ to $\nu = 1.3(1)$. However, the flow to the insulating phase is slow and we cannot be sure these exponents are really distinct; in any case they are very far from the random value $\nu \approx 2.4$. Moreover, the multifractal spectra in the quasiperiodic case also seem nonuniversal. Determining these exponents more accurately, and identifying whether they are stable with respect to the Harris bound $\nu = 1$, are interesting questions for future work. The irregular dependence of the phase diagram and exponents on the quasiperiodic angle $\theta$ and the quasienergy $\omega$---as well as the sensitivity of the exponents to adding additional modulations---suggests that these quantities are sensitive to high-order scattering processes that depend on the precise kinematics of the quasiperiodic potential. It remains an open challenge to develop an analytic framework for understanding this dependence. It would also be interesting to develop a scaling theory of transport in the critical phase, and to look for similar critical phases in quasiperiodic systems in other Altland-Zirnbauer symmetry classes~\cite{altland1997nonstandard}.


\vspace{.3cm}
\section*{Acknowledgement} We acknowlewdge discussions with Ilya Gruzberg. Jonas Karcher is supported by the Army Research Office under the MURI program, grant number W911NF-22-2-0103.  We acknowledge support from NSF DMR-2103938 (S.G.), DMR-2104141 (R.V.) and the Alfred P. Sloan Foundation through Sloan Research Fellowships (R.V.) .

\bibliography{QP}

\onecolumngrid
\appendix

\section{Multifractal spectra in the metallic phase and on the novel critical line}
\label{app:multi}
Here, we extend the analysis of the multifractal spectrum presented in Fig.~\ref{fig:mf_fss}. The multifractal spectrum is defined by the scaling of the $q$-th participation ratio:
\begin{align}
	\mathcal{P}_q\equiv N_B^{-1}\left\langle \sum_i (\mu_i)^q \right\rangle \sim (b/L)^{2+x_q}.
\end{align}
The anomalous dimensions $x_q$ encode the information about the fractal dimensions $f(\alpha)$ of the sets where the wavefunction scales as $L^{-\alpha}$, more precisely $(q,x_q)$ is related to $(\alpha, f(\alpha))$ by a Legendre transform \cite{Evers2008}. Typically the spectrum is approximately parabolic $x_q\approx \beta q(1-q) \equiv \beta z_q$. The Weyl symmetry relation $x_q=x_{1-q}$ constrains random class A Anderson transition MF spectra  \cite{Evers2008}.

In Fig.~\ref{fig:_mf}, we show multifractal spectra $x_{q}$ for various quasiperiodic angles $\theta$. Black dots are data points from a fit of the exponent over system sizes $L=96\ldots 600$. We show data for a point in the phase diagram close to the boundary of the extended critical phase for $\theta=\pi/6.0$ and deep in the metallic phase for $\theta=\pi/6.2$ and $\ln d^{-1}=0.0$. These behave strikingly similar. Finally we show data for the novel critical point for $\theta=\pi/6.2$ and $d=0.11 \approx d_c$. There is strong multifractality and the multifractal spectrum is still approximately parabolic.
In particular the transition at $\theta=\pi/6.2$ displays strong violations of the symmetry relation $x_q=x_{1-q}$ that holds exactly in the uncorrelated random case.

\begin{figure*}
	\centering
	\includegraphics[width = .29\textwidth]{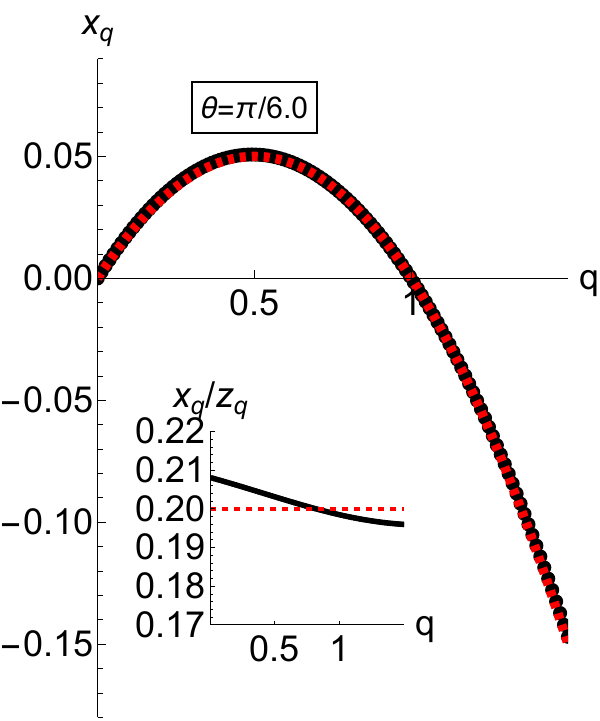}
	\includegraphics[width = .29\textwidth]{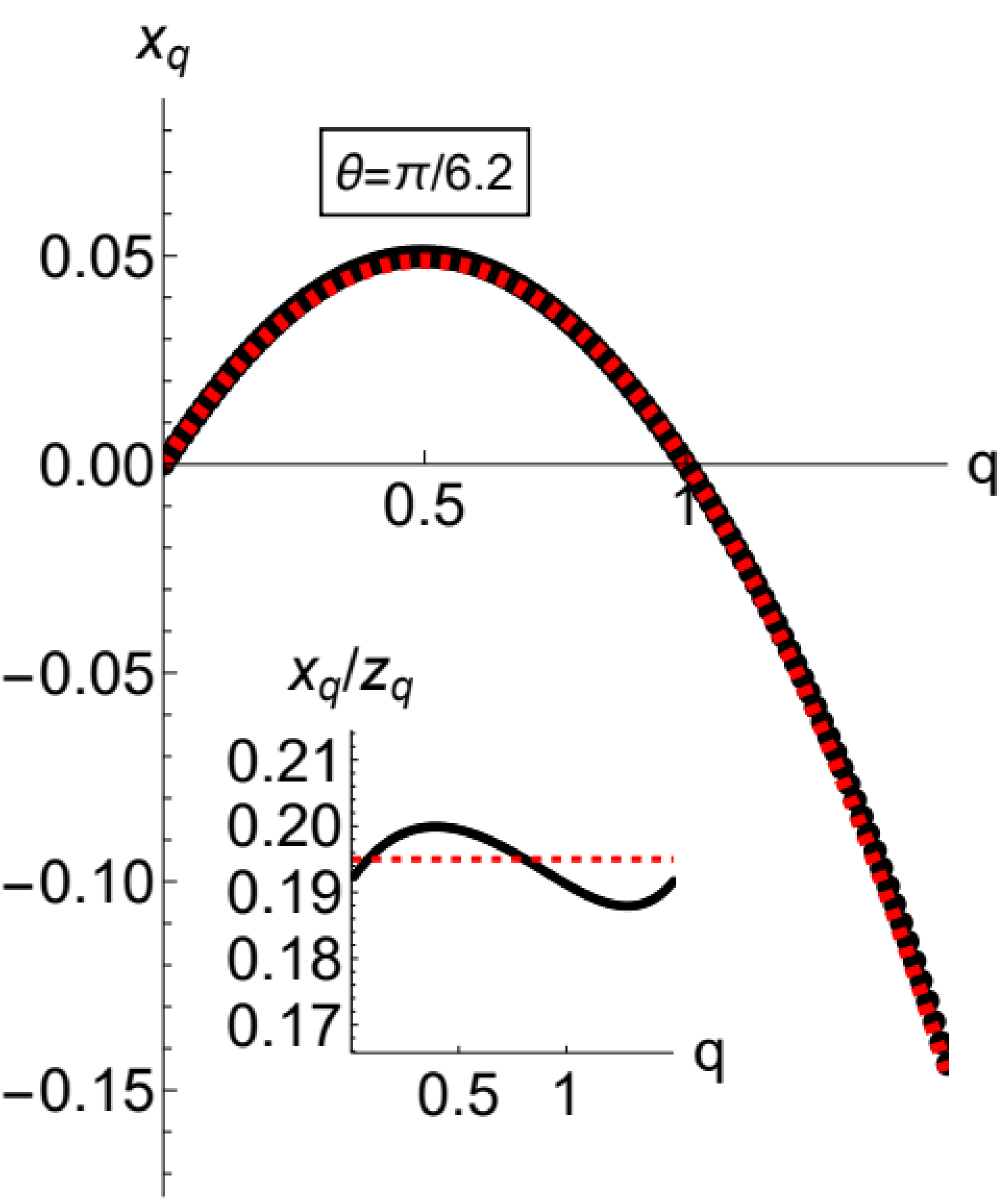}
	\includegraphics[width = .29\textwidth]{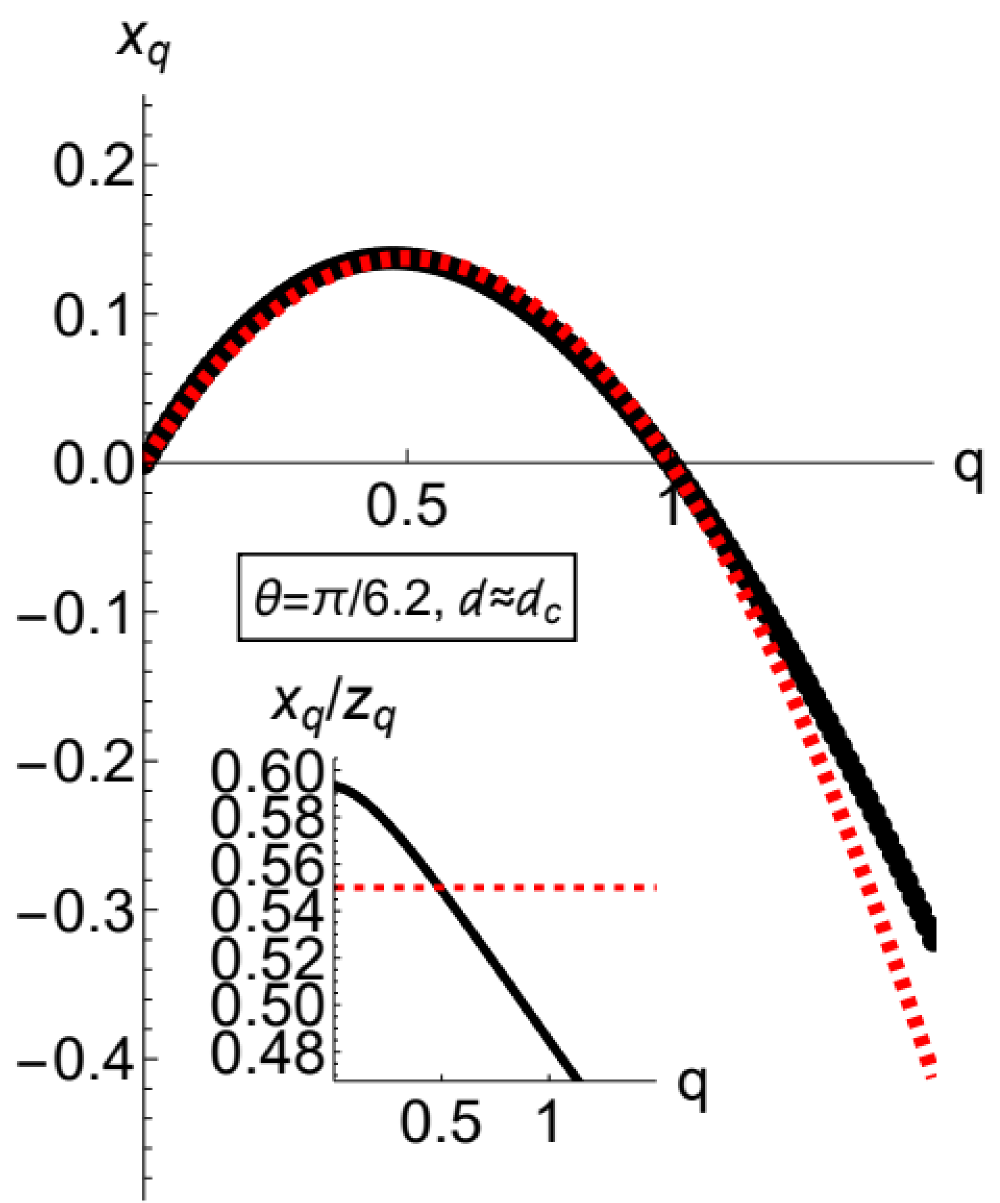}
	\caption{{\bf Multifractal spectra $x_{q}$ for various quasiperiodic angles $\theta$.} Black dots are data points from a fit of the exponent over system sizes $L=96\ldots 600$. \textit{Left panel:} close to the boundary of the extended critical phase for $\theta=\pi/6.0$ \textit{Middle panel:} deep in the metallic phase $\theta=\pi/6.2$ and $d=0.0$. The result is very similar to the data shown in the right panel. \textit{Right panel:} novel critical point for $\theta=\pi/6.2$ and $d=0.11 \approx d_c$. There is strong multifractality and the multifractal spectrum is still approximately parabolic.}
	\label{fig:_mf}
\end{figure*}

\section{Crossover to randomness for many tones: QP CC transfer matrix}
\label{app:transfer}

In this appendix, we demonstrate the crossover from the quasiperiodic to the random fixed point as tones are added in the quasiperiodic potential. We do so using transfer matrix studies of the U(1) Chalker-Coddigton network (CCN) in class A.

We provide additional numerical data to support the conjecture on the crossover between randomness and quasiperiodic fixed points (see Fig.~\ref{fig:crossover_iqh}). We perform a transfer matrix analysis of a quasi-1D (Q1D) $L\times W$ strip of the system with $L\gg W$. The observable we study is the second Lyapunov exponent $\xi_W$. In the limit $W\rightarrow \infty$, the ratio $\xi_W/W$ is an observable suitable for finite size scaling near criticality.

We can write the network model scattering matrix (see Fig.~\ref{fig:scat}, colors are chosen to match link colors there) as: 
\begin{align}
	\begin{pmatrix}
		c^{-1}e^{i\epsilon}{e^{-i\phi_{t+1,u}}} &
		sc^{-1} {e^{-i\phi_{t+1,u}}} {e^{i\phi_{t,u+1}}} \\
		s c^{-1}  
		& (c+s^2 c^{-1} ) e^{-i\epsilon}{e^{i\phi_{t,u+1}}}
	\end{pmatrix}
	\begin{pmatrix}
		\textcolor{black}{\ell_{t,u}} \\
		{\ell_{t,u+1}}
	\end{pmatrix}
	&= 
	\begin{pmatrix}
		{\ell_{t+1,u}} \\
		{\ell_{t+1,u+1}}
	\end{pmatrix}, \nonumber\\ 
	\begin{pmatrix}
		s^{-1} e^{i\epsilon} \textcolor{black}{e^{-i\phi_{t+2,u}}} & -s^{-1}c\textcolor{black}{e^{-i\phi_{t+2,u}}}{e^{i\phi_{t+1,u-1}}}\\
		-s^{-1}c & (s +s^{-1}c^2)e^{-i\epsilon} {e^{i\phi_{t+1,u-1}}}
	\end{pmatrix}
	\begin{pmatrix}
		{\ell_{t+1,u}}\\
		{\ell_{t+1,u-1}}
	\end{pmatrix}
	&=
	\begin{pmatrix}
		\textcolor{black}{\ell_{t+2,u}}\\
		{\ell_{t+2,u-1}}
	\end{pmatrix}
\end{align}
in order to find the transfer matrix of the system relating the link amplitudes $\ell_{t+1,u}$ of slice $t$ to those of slice $t+1$. Using standard methods \cite{kramer2005random}, we can find all Lyapunov exponents of a Q1D strip.

We interpolate between quasiperiodic link modulations with two different $\theta$ using a parameter $w\in [0,1]$:
\begin{align}
	\phi(r) &= w \phi_1(r) + (1-w)\phi_2(r)
	& \phi_k(r) &= 2\pi\sum_{i=1,2} \cos\left(\sum_jA_{ij}^{(k)}r_j + \gamma_i\right), \nonumber\\
	A^{(k)}_{11} &= A^{(k)}_{22} = \varphi\cos\theta_k, A^{(k)}_{12} = - A^{(k)}_{21} = \varphi\sin\theta_k. \label{eq:qp2}
\end{align}
At $w=0$ (or $w=1$) only one tone is present and at $w=0.5$ they are perfectly mixed. This way, we can see the random/QP crossover for different QP parameters. For the numerical calculations, we choose $W=12$, 16, 24, 32, 48, 64, 80, 96, 112, 128, 144, 160, 176, 192, 208, 224, 240, 256, 272, 288, 304, 320, 336, 352, 368, 384 and $L=10^5$. In Fig.~\ref{fig:crossover}, we show additional supporting data on the crossover between randomness and quasiperiodic fixed points. It does not seem to occur when the second modulation is sufficiently weak, instead $\xi_W/W$ shows a discontinous jump at a finite value $1>w_c>0$ of the interpolation. This effect seems to be generic as it appears for various pairs of $\theta_1, \theta_2$.

\begin{figure}
	\centering
	\includegraphics[width = .45\textwidth]{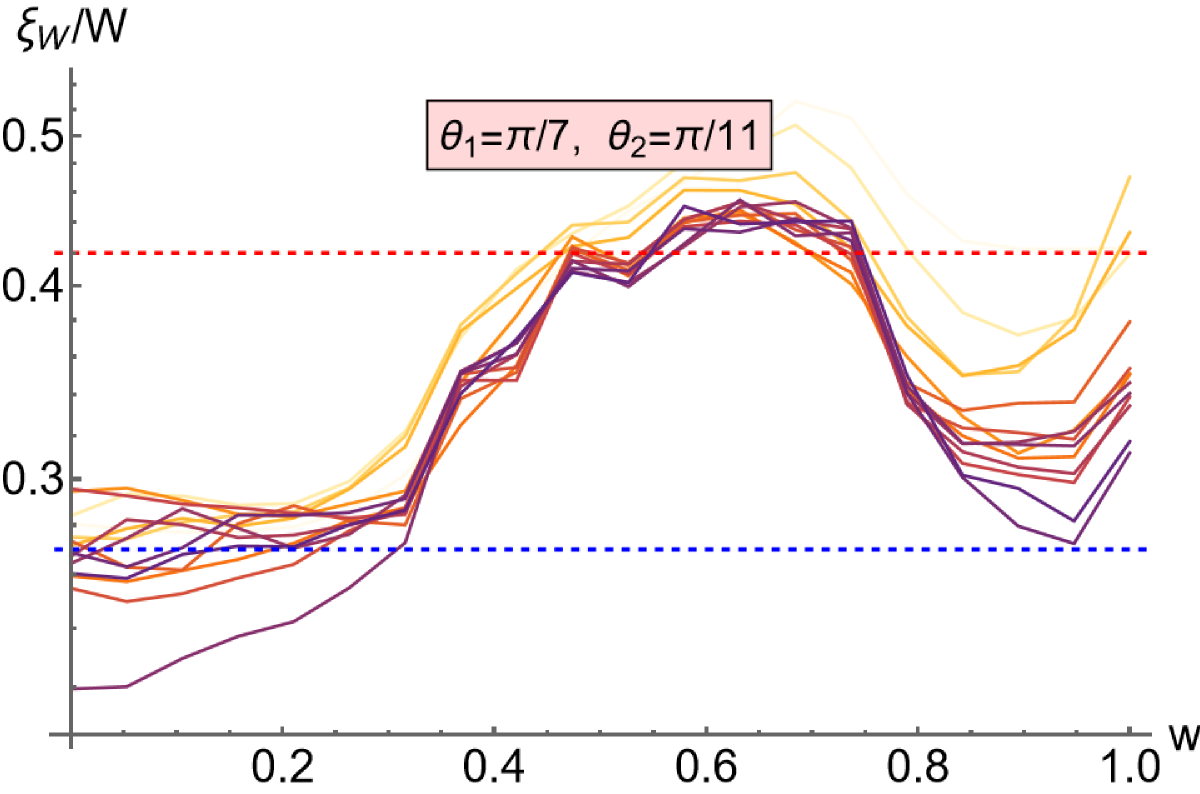}
	\includegraphics[width = .45\textwidth]{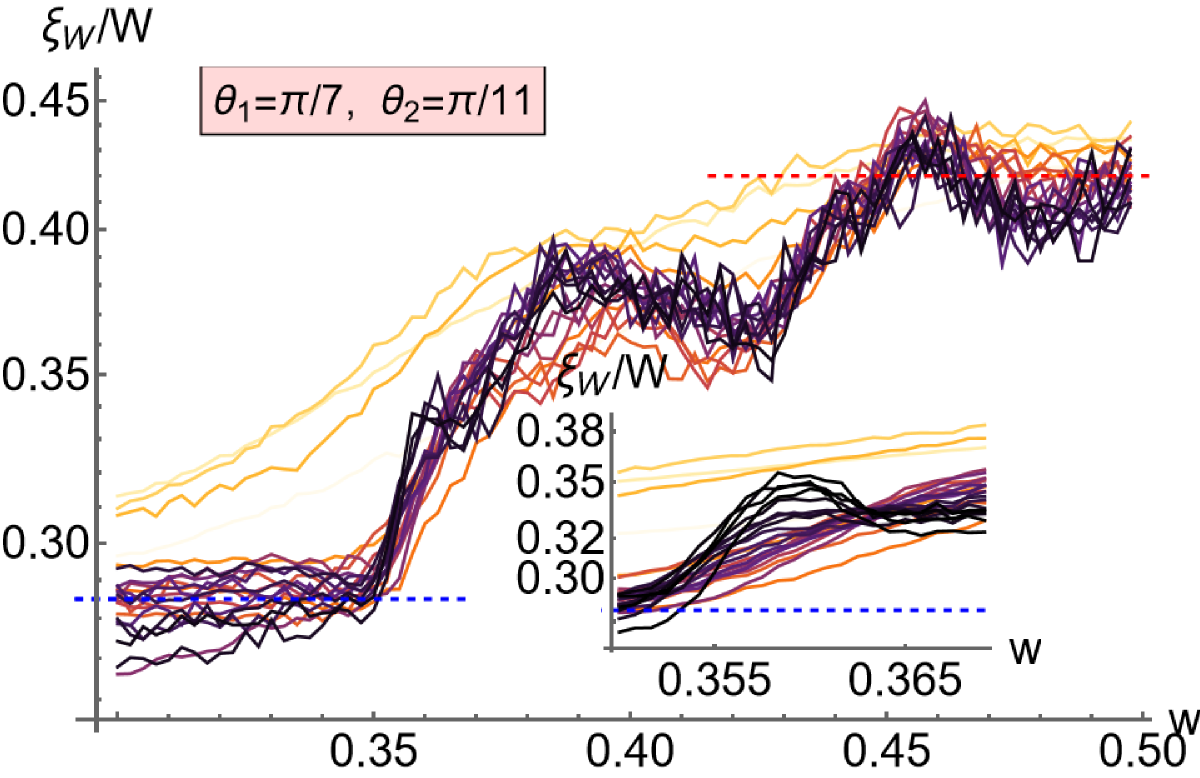}
	\includegraphics[width = .45\textwidth]{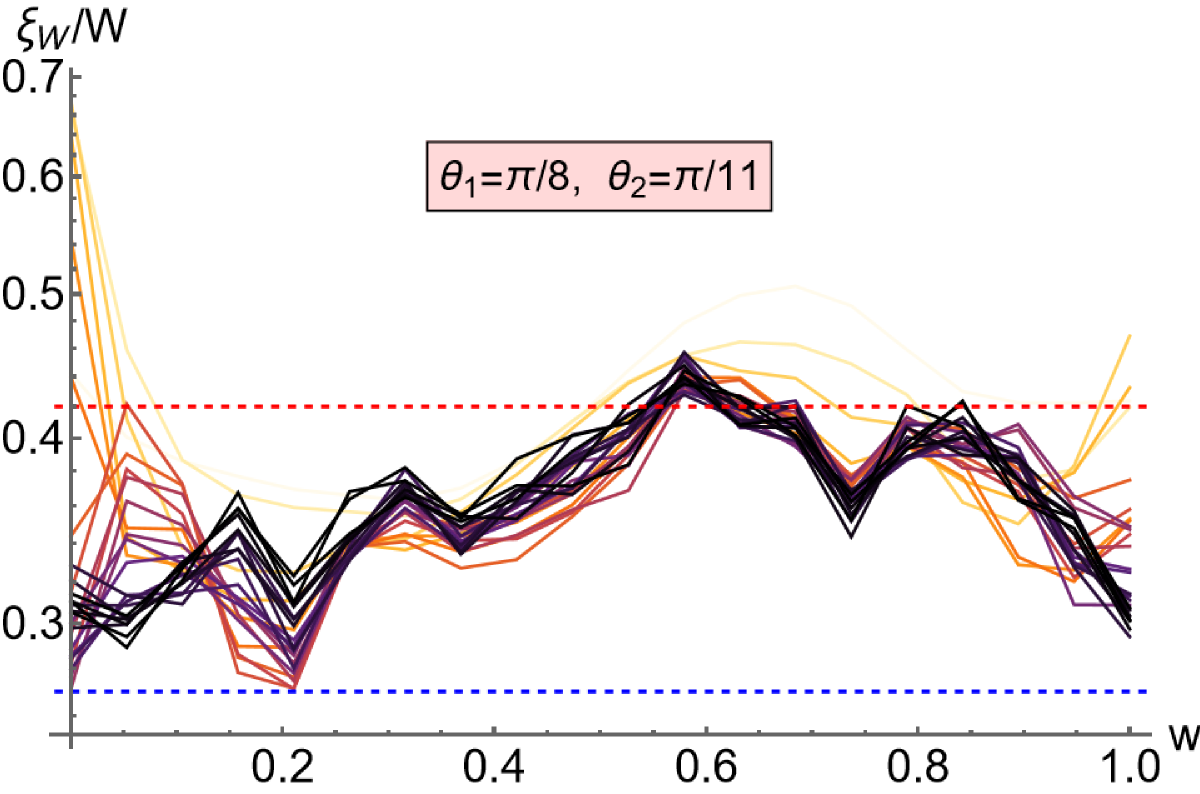}
	\includegraphics[width = .45\textwidth]{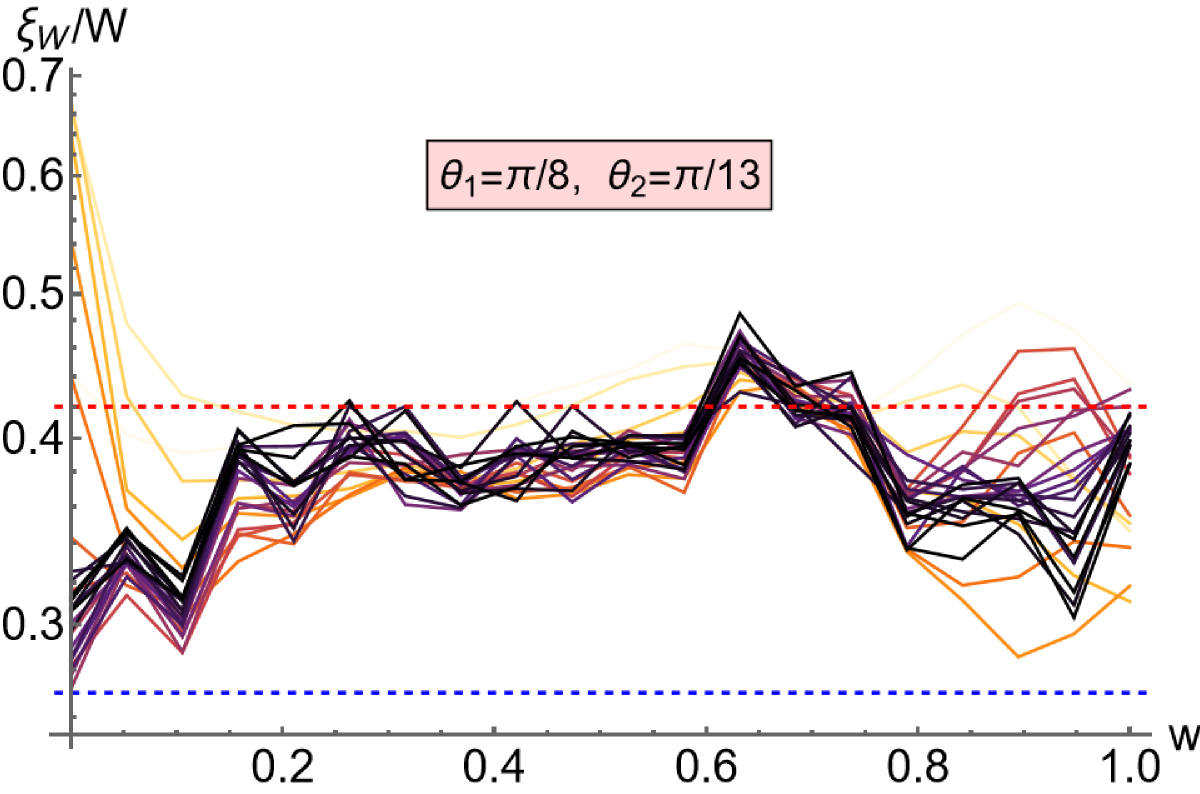}
	\includegraphics[width = .45\textwidth]{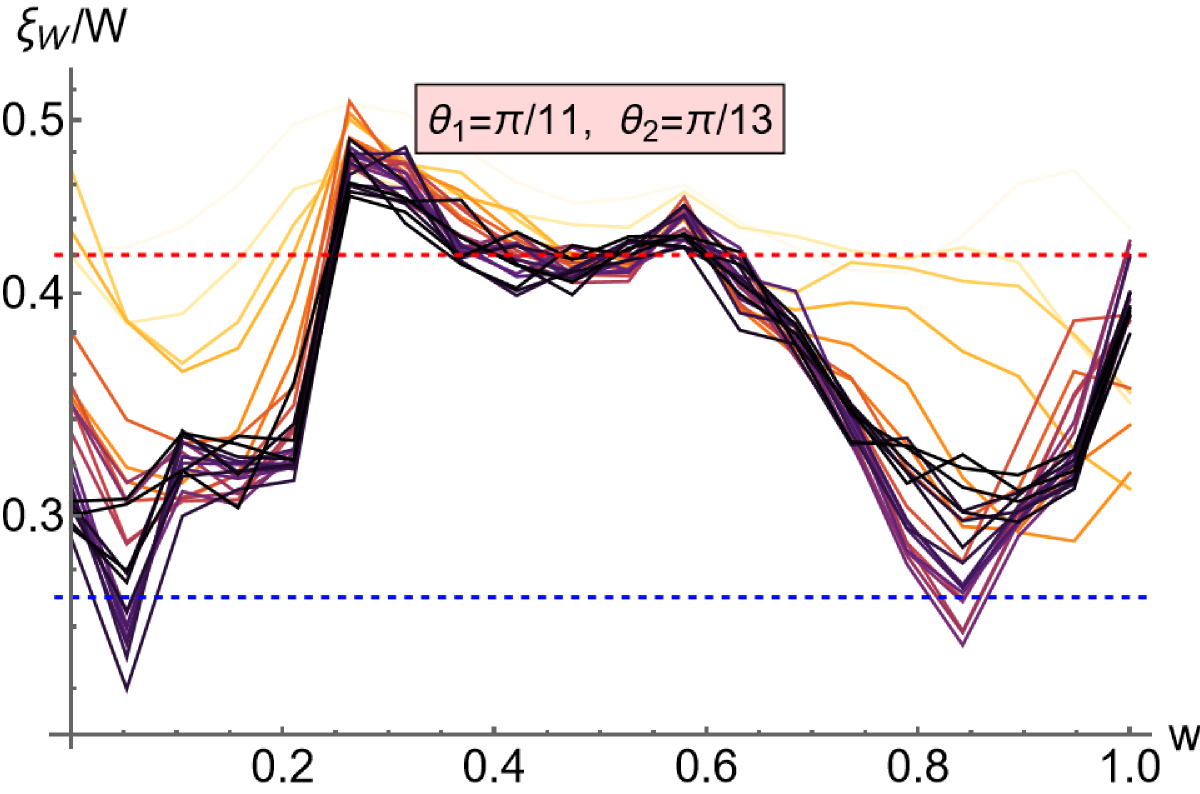}
	\includegraphics[width = .45\textwidth]{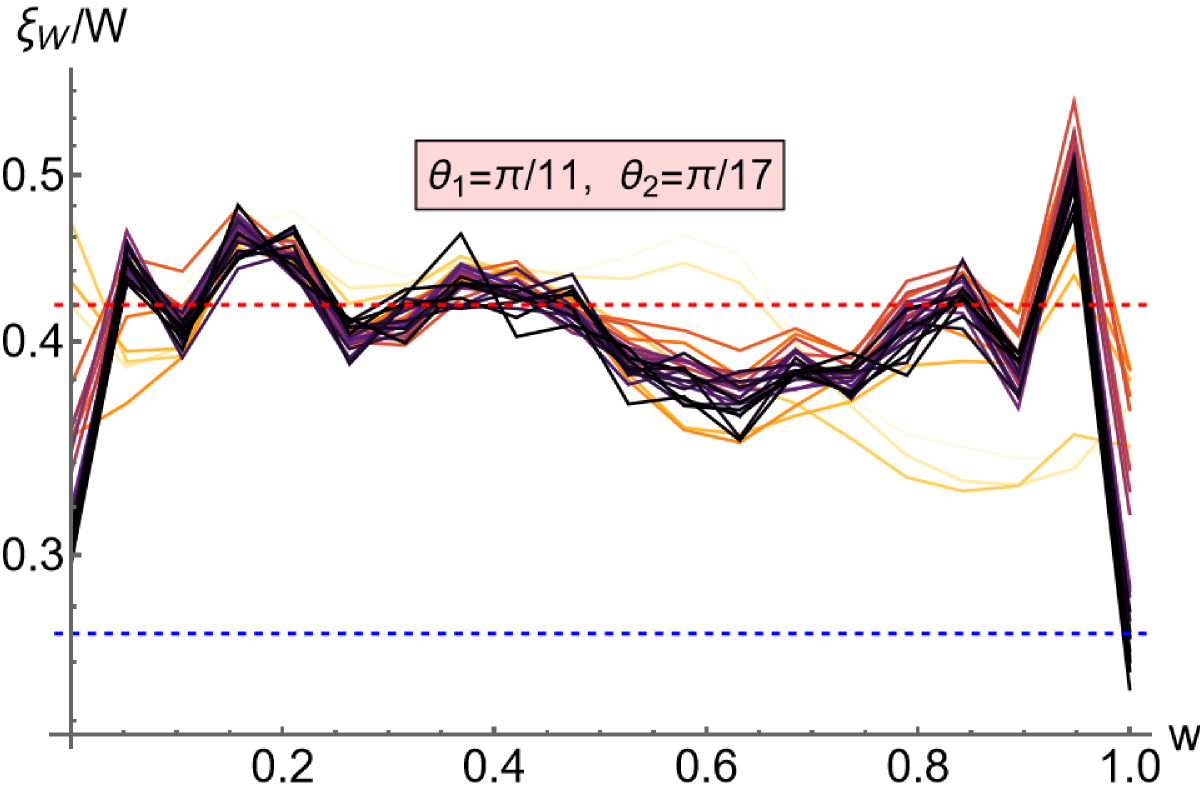}
	\caption{{\bf Crossover between randomness and quasiperiodic fixed points. } The observable we study is the second Lyapunov exponent $\xi_W$ of a quasi-1D $L\times W$ strip of the system. We interpolate between different quasiperiodic link modulations with a parameter $w$. At $w=0$ (or $1$) only one (or the other) tone is present and at $w=0.5$ they are evenly mixed. Dashed red and blue lines are guids to the eye for QP and random value of $\xi_W$. }
	\label{fig:crossover}
\end{figure}

\end{document}